\documentclass{ptapap}

\author{L\'aszl\'o Moln\'ar}[KONKOLY]
\affil[KONKOLY]{Konkoly Observatory, Research Centre for Astronomy and Earth Sciences,\\ Konkoly Thege Mikl\'os \'ut 15-17, H-1121 Budapest, Hungary}

\title{RR~Lyraes and Cepheids: the Photometric Revolution from Space}
\headtitle{Space-based Photometry Revolution}
\begin{document}

\maketitle

\begin{abstract}
Continuous, high-precision photometry from space revolutionized many fields of stellar astrophysics, and that extends to the well-studied families of RR~Lyrae and Cepheid variable stars as well. After the pioneering work of \textit{MOST}, the \textit{CoRoT} and \textit{Kepler} missions released an avalanche of discoveries. We found signals that needed exquisite precision, such as an abundance of additional modes and granulation. Other discoveries, like period doubling, simply needed us to break away from the day-night cycle of the Earth. And the future holds more possibilities, with the \textit{BRITE}, \textit{K2}, and \textit{Gaia} missions at full swing; \textit{TESS}, taking physical shape; and \textit{PLATO} securing mission adoption. Here I summarize some of these discoveries and the expectations from future missions.
\end{abstract}

\section{From the Beginnings to the \textit{Kepler} Era}

Space-based photometry, in general, includes the majority of observations made by space telescopes, for a wide variety of reasons. One broad application focuses on the temporal variation of the measured fluxes, via time series photometry. The collection of time series data still has a wide usage, and in some cases, the nature of the flux variation itself takes a back seat to help solve different questions. A prominent example is the numerous extragalactic Cepheid light curves obtained with the Hubble Space Telescope that were used to pin down the value of $H_0$, the Hubble constant \citep{riess2016}. Here we shall discuss missions that study stars via space-based, (quasi-)continuous time-series photometry, with emphasis on some particular subgroup of stars: the most classical of classical pulsators, RR~Lyrae and Cepheid stars. 

\paragraph{The missions.} The story of space-based, continuous time-series photometry of stellar targets goes back a long way. One early prediction was that the eventual discovery of small, rocky planets around other stars through their transits would need photometric precision that is only achievable from space, to avoid the effects caused by the atmosphere \citep{borucki1984}. These initial findings eventually developed into a proposed NASA mission called \textit{FRESIP} (FRequency of Earth-size Inner Planets), but later renamed to \textit{Kepler} \citep{borucki2016}. 

Meanwhile, other missions were also in development, to pursue slightly different goals. Photometric stability and extended observations that circumvent the night-day cycles of ground-based observations are a necessity not only for transits but for stellar astrophysics too. Solar-like oscillations, in particular, were long sought-after, promising to extend the toolkit of helioseismology to stars other than the Sun. Even before the \textit{FRESIP/Kepler} proposals, European researchers were already pursuing to fly an instrument dedicated to stellar activity and variability as a standalone mission or as a secondary payload \citep{roxburgh2006}. These efforts culminated in two projects that left the Earth. One was \textit{EVRIS} (Etude de la Variabilit\'e, de la Rotation et des Int\'erieurs Stellaires), a 9\,cm telescope that was supposed to observe bright stars on board the Russian \textit{Mars--96} spacecraft en route to Mars \citep{baglin1991}. Unfortunately, the mission never left low-Earth orbit and burned up in the atmosphere. The first successful European space-photometric mission came a decade later, with the launch of \textit{CoRoT} (Convection, Rotation and planetary Transits), a 24\,cm telescope in polar Earth orbit that executed a program shared between asteroseismology and exoplanets until 2012 \citep{baglin2002}.

In the US and Canada, while successive \textit{Kepler} proposals got rejected, other photometric missions emerged. The Canadian \textit{MOST} (Microvariability and Oscillations of Stars), a small satellite with a 15\,cm telescope was launched in 2003 to execute a one-year asteroseismic mission \citep{walker2003}. The sturdy satellite is still functional at the time of writing, 14 years later. In the US, the malfunction of the main instrument of the \textit{WIRE} (Wide-field Infrared Explorer) space telescope in 1999 led to the unexpected possibility to use the 5\,cm optical star tracker telescope for asteroseismology for several years \citep{buzasi2002}. Further contributions to the field include the \textit{SMEI} (Solar Mass Ejection Imager) all-sky camera on the \textit{Coriolis} satellite that observed between 2003--11 \citep{hick2007}, and the High-Resolution Imager on the \textit{Deep Impact} space probe that was utilized to measure transiting exoplanets in 2008 \citep{ballard2011,christiansen2011}. 

And then \textit{Kepler} was launched in 2009. But let us first summarize the discoveries made by the missions before that. 

\begin{figure}
\includegraphics[width=\textwidth]{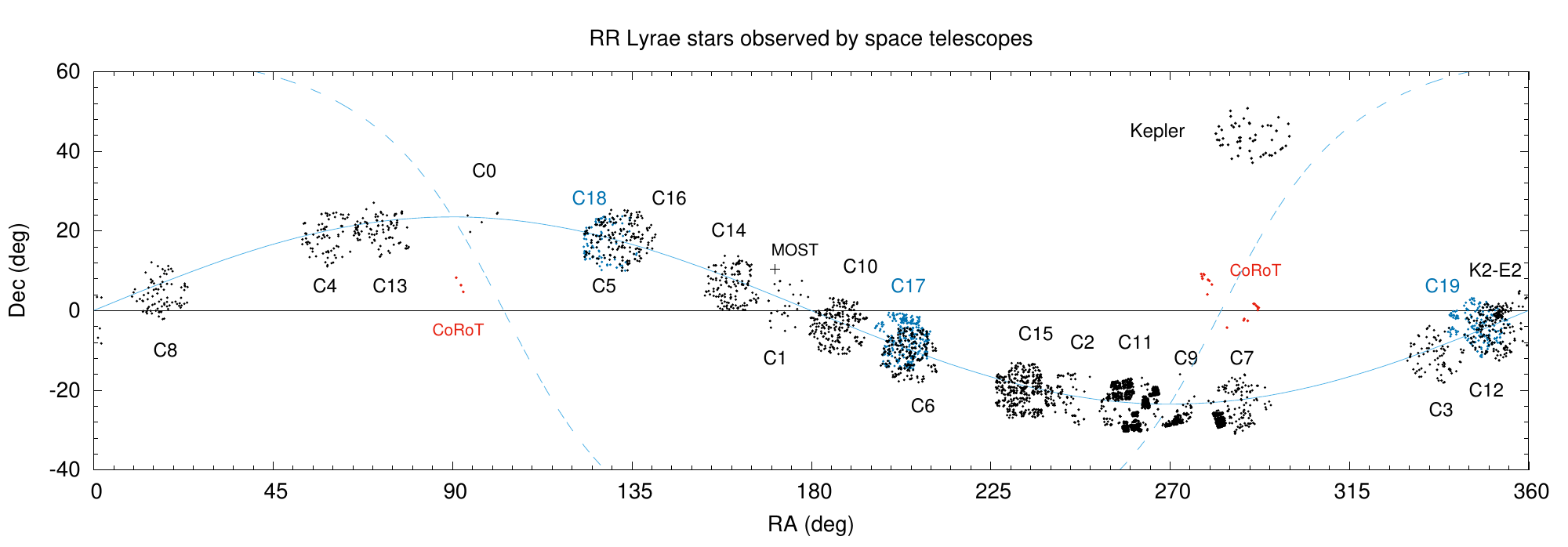}
\caption{RR~Lyrae stars observed by various space missions. Black and blue points show targeted and proposed \textit{Kepler} and \textit{K2} observations, red points are \textit{CoRoT} targets, and cross marks the single \textit{MOST} target.}
\label{fig:rrlmap}
\end{figure}

\paragraph{The first discoveries.} The missions until \textit{Kepler} were limited to a few weeks of observations (extending to $150$\thinspace d with the long runs of \textit{CoRoT}), and to relatively bright targets. Nevertheless, the combination of stable, precision photometry with extended coverage, led to discoveries in RR~Lyrae and Cepheid stars. While these stars have been studied extensively for more than a century now, they still hold mysteries and convey valuable information for both stellar astrophysics and near-field cosmology. The first RR~Lyrae, AQ~Leo, a representative of the RRd (double-mode) subclass was observed from space by \textit{MOST}. \citet{gruberbauer} discovered low-amplitude additional signals in the star that were attributed to non-radial modes, the first clear detection of such additional modes in RR~Lyrae stars. 

In parallel, \citet{bruntt2008} combined the extended, but low-precision \textit{SMEI} data with the shorter, but much more dense and precise \textit{WIRE} observations, and with additional radial-velocity data for the Cepheid star Polaris ($\alpha$~UMi). The observations showed that the pulsation is, in fact, not switching off in the star, as it was assumed before, but instead its amplitude started to increase after reaching minimum in the years around 2000. They also detected some small power excess in the low-frequency end of the \textit{WIRE} data that they tentatively interpreted as granulation noise in the star.

Interestingly, the pioneering observations were not followed up with other targets by those missions. Instead, the thread was picked up by the \textit{CoRoT} mission that started to accumulate longer light curves for multiple objects. With 143 days of continuous observations collected for the short-period, fundamental-mode star V1127 Aql, we had the best picture of the Blazhko effect \citep{v1127}. The star also showed various additional modes, whose origins were quite mysterious at the time. Recently, similar signals were found in multiple short-period RRab stars among the OGLE Bulge stars \citep{prudil2017}. 

Variations or multiple periodicities in the Blazhko effect, the amplitude and phase modulation of the pulsation, were long suspected \citep[see, e.g.,][]{sodor2006}. Another RR~Lyrae, CoRoT\,105288363, showed us that consecutive Blazhko cycles can, indeed, have different shapes and amplitudes \citep{guggenberger2011}.

\begin{figure}
\includegraphics[width=\textwidth]{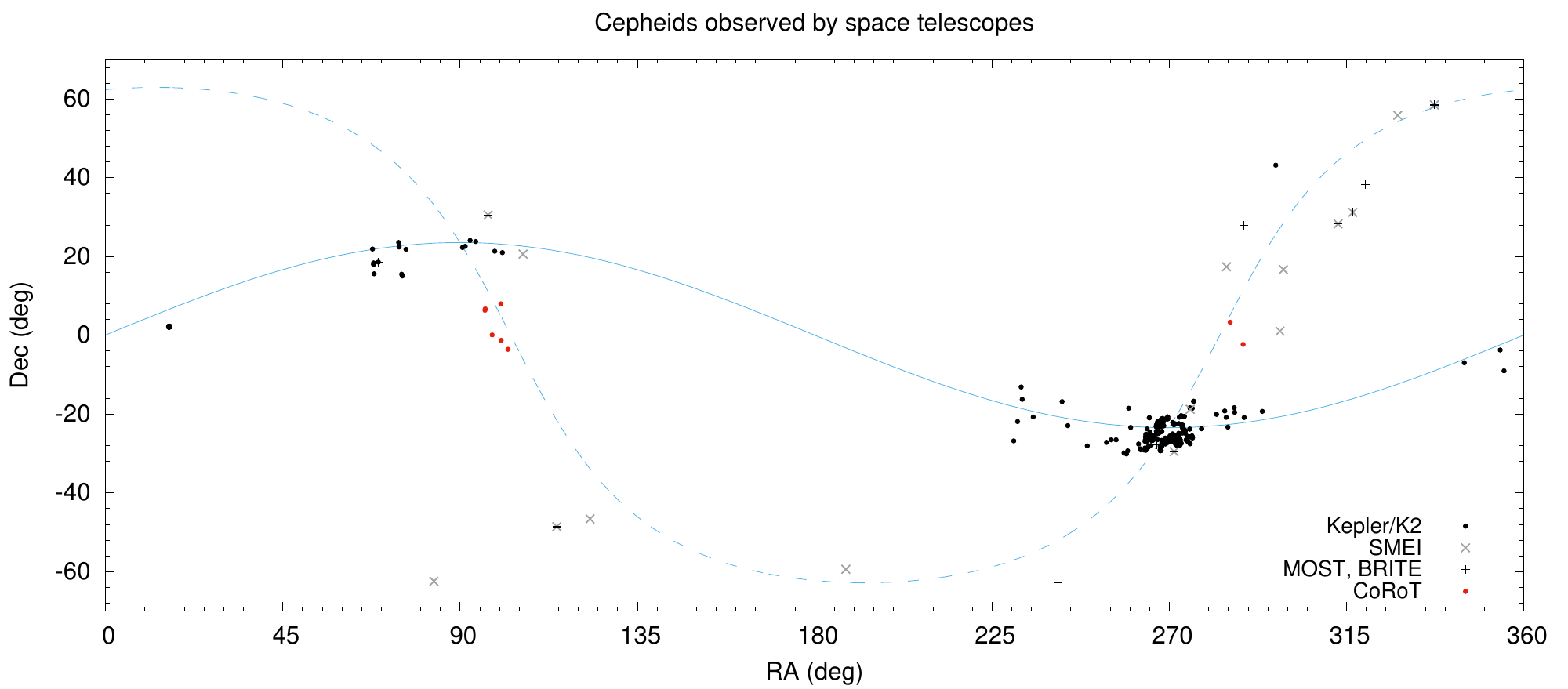}
\caption{Cepheid stars observed by various space missions. Notation is similar to Fig.~\ref{fig:rrlmap}. The only target missing from this map is Polaris.}
\label{fig:cepmap}
\end{figure}

\section{Enter \textit{Kepler}}

While these missions provided a new look into variable stars, the field has been revolutionised by the \textit{Kepler} mission. A large, 95\,cm aperture telescope, with a huge field-of-view spanning about 100 square degrees, in an Earth-trailing orbit far from the planet, and an observing mode that provided multi-year coverage, made \textit{Kepler} vastly superior to its predecessors. And the mission was put to good use. It gathered 4 years of continuous observations in a very wide brightness range that will not be surpassed for many years to come. But then, in 2013, the telescope was left with one reaction wheel short to continue its mission. Soon an ingenious plan called the \textit{K2} mission was devised to reorient \textit{Kepler} towards the Ecliptic to follow a step-and-stare approach and to look at a new field of view at about every three months \citep{k2}. It is now clear that this massive hack to save a crippled space telescope has evolved into a hugely successful and versatile astrophysical mission.

\paragraph{The power of continuous data.} The first \textit{Kepler} light curves of RR~Lyrae stars were simply shocking: multiple stars, including RR~Lyr itself, showed clearly alternating cycle amplitudes in the pulsation \citep{szabo2010,kolenberg2011}. These variations, called period doubling (since the repeating pattern is now two cycles long) should have been clearly detectable from the ground. That they were not can be traced back to a simple cause: the typical half-day-long pulsation periods of these stars make it virtually impossible to observe consecutive cycles with the same telescope. 

Nevertheless, period doubling was not the only new signal found in RR~Lyraes. The \textit{Kepler} data revealed additional frequency components below the millimagnitude level in fundamental-mode stars \citep{benko2010,benko2014,molnar2012}. While most of these signals can be grouped into three main categories (near the expected ranges of first and second overtones, plus the half-integer peaks of period doubling), they spread out in the Petersen-diagram, indicating that many of them must be non-radial modes \citep{molnar2017}. Interestingly, all of these modes were identified in modulated stars, while all non-Blazhko RRab stars appeared to be monoperiodic, down to sub-millimag amplitudes, suggesting some connection between the two phenomena. However, it is not always easy to separate modulated and non-modulated stars from each other. While the presence of amplitude modulation is obvious in most \textit{Kepler} Blazhko targets, some stars needed careful processing and analysis to detect small changes in the light curves \citep{nemec2011,benkoszabo}. In these cases, phase variation and changes in the relative Fourier coefficients are better indicators for modulation than the pulsation amplitude itself. The light curves also revealed that the majority of modulated stars have either multiperiodic or changing modulation cycles \citep{benko2014}. \citet{plachy2014} investigated if the modulation could be chaotic, but even the 4-year-long \textit{Kepler} data covered too few modulation cycles for a conclusive answer. 

Analysis of the four known RRc stars led to similar discoveries. All four stars show various additional modes, including the so-called $f_{\rm x}$ mode or $0.61$-mode that has been detected in a large number of RRc and RRd stars, both from space and from the ground \citep{moskalik2015,netzel2015a}. The first additional frequencies found in AQ~Leo also fit into this pattern. Similar signals were found in overtone Cepheids as well, highlighting the peculiarity of these modes and their connection to the first overtone \citep{moskalik2009}. Yet another group of modes in RRc stars fall below the (expected) frequency of the fundamental mode \citep{netzel2015b}.

The \textit{Kepler} field-of-view included only a single Cepheid, V1154~Cyg, but 4 years of continuous observations revealed even the smallest details about this fundamental-mode star. While the pulsation period was found to be stable in the long term, fluctuations in the cycle shape and length were clearly detectable \citep{szabo2011,derekas2012}. Moreover, a low-level modulation is also present in the pulsation, on top of the fluctuations. Then a further signal was found, hiding behind the pulsation: the first clear detection of granulation noise in a Cepheid. It was even more intriguing that solar-like oscillations were not seen in the star, suggesting that they are either strongly suppressed, or absent altogether \citep{derekas2017}. 

\paragraph{Advances in understanding.} These discoveries have presented new challenges to our understanding of stellar pulsation. Model calculations soon revealed that period doubling is caused by non-linear interactions between pulsation modes. The 9th radial overtone may become trapped between the surface and the partial ionization zone and turn into a so-called strange mode. This mode then can lock into a strong resonance with the fundamental mode with a period ratio of $P_0/P_9 = 9/2$ \citep{kmsz11}. When that happens, we are not simply observing two modes anymore, but instead, the fundamental mode bifurcates into a period-doubled limit cycle, repeating two pulsation cycles with different amplitudes. These resonances could potentially lead to modulation in some calculations, suggesting that mode resonances and interactions could be behind the Blazhko effect \citep{bk11}. We emphasize here that the mode resonance hypothesis is currently the only model that is not in contradiction with the observations. 

Further calculations revealed that the period-doubled fundamental mode can become unstable against the first overtone too, outside the classical RRd regime. The appearance of a low-amplitude first overtone in RRab stars could explain at least some observations of additional modes \citep{molnar2012,kollath2016}. Detailed analysis of these three-mode hydrodynamic models showed that they can become chaotic, an unexpected result from stellar models with relatively low mode growth rates \citep{plachy2013}. 

The presence of other additional modes has been also intriguing, especially the abundance of $f_{\rm x}$-type modes in overtone stars. There is no plausible explanation in our current framework for their appearance. However, \citet{dziem} suggests that we actually observed a harmonic of non-radial $\ell = 8, 9$ modes (also $\ell = 7$ for Cepheids), and the intrinsic pulsation frequency at $f_{\rm x}/2$ is observed, in the majority of cases, with smaller amplitudes due to cancellation effects. 

It has been long known that the shape of RR~Lyrae light curves is affected by their chemical composition. The photometric metallicity relation has been calibrated for the \textit{Kepler} passband based on high-resolution spectra \citep{nemec2013}. This semi-empirical relation is accurate to about $\pm0.1$~dex in [Fe/H], except for stars with the strongest modulation. However, it is important to keep in mind that Fourier parameters can change drastically in some Blazhko stars, which could affect the derived photometric metallicities too (B\'odi et al., these proceedings).

\section{After \textit{Kepler}}

The \textit{Kepler} mission has transformed stellar astrophysics, but it was nevertheless limited to a single patch of sky, with a limited stellar sample. Therefore, other missions were utilized in new ways, either by mining their archives or by proposing new observations. The \textit{CoRoT} light curves also revealed period doubling and additional modes in multiple RR~Lyrae stars \citep{szabo2014,benko2016}. No significant period fluctuations were found in the Cepheids observed by the mission, but one double-mode star exhibited the same, low-frequency additional mode that was detected in RRc stars too \citep{poretti2014,poretti2015}.

Many Cepheids are bright enough to be detected even by the smallest-aperture missions. The \textit{BRITE} fleet of satellites observed multiple targets, but unfortunately only those gathered with the red-filter satellites turned out to be useful. Possible detection of modulation and additional modes in some targets were reported by \citet{smolec2017}. The venerable \textit{MOST} satellite also observed some Cepheids. It detected slight differences between the stability of pulsation in fundamental-mode and overtone stars as well as period doubling in the peculiar, modulated Cepheid, V473 Lyr \citep{evans2015,molnar2017b}.

Meanwhile, the saga of \textit{Kepler} itself was far from over. \textit{K2}, the new mission, opened up many new fields for observation, from the halo to the Sagittarius stream and the Galactic Bulge. Although campaigns are limited to 70--80 days, the fields amassed thousands of RR~Lyrae and hundreds of Cepheid observations (Figs~\ref{fig:rrlmap}--\ref{fig:cepmap}, \citealt{plachy2016}). These included lesser-studied subtypes like RRd stars, including a modulated one, as well as variables in globular clusters and nearby galaxies, but the processing of the light curves is still ongoing \citep[see, e.g.,][]{molnar2015,kurtz2016,kuehn2017,plachy2017a}. Type II Cepheids of the W~Vir subclass showed clear cycle-to-cycle variations \citep{plachy2017b}. 

The \textit{K2} goldmine will serve us with wonderful discoveries for years to come, but we cannot rest: new missions will provide even more data. The European \textit{Gaia} spacecraft does not fit into our description of continuous photometry, but it is nevertheless collecting sparse observations for the entire sky. Many new variable stars are expected to come from the photometry and the extremely precise astrometry will undoubtedly mean another revolution for stellar astronomy \citep{clementini2016}. And then, from 2018, the American \textit{TESS} mission will produce proper light curves for almost the whole sky -- although not quite to the same depth as \textit{Kepler} and \textit{Gaia} may reach \citep{ricker2014}. And with all these missions at full swing, the photometric revolution may enter a new phase, where synergies between missions, such as combination of photometry with astrometry, or revisits by successive missions and spectroscopic follow-up will become just as important as the light curves themselves.

\acknowledgements{The research leading to these results have been supported by the Hungarian National Research, Development and Innovation Office (NKFIH) grant PD-116175 and the Lend\"ulet LP2014-17 grant of the Hungarian Academy of Sciences. L.~M.\ was supported by the J\'anos Bolyai Research Scholarship of the Hungarian Academy of Sciences. Funding for the \textit{Kepler} and \textit{K2} missions are provided by the NASA Science Mission Directorate. I acknowledge with thanks the hospitality of the organizers of the RR~Lyrae 2017 meeting in Niepo\l{}omice, Poland.}

\bibliographystyle{ptapap}
\bibliography{molnar}

\begin{thebibliography}{55}
\providecommand{\natexlab}[1]{#1}
\providecommand{\url}[1]{\texttt{#1}}
\providecommand{\urlprefix}{URL }
\providecommand{\eprint}[2][]{\url{#2}}

\bibitem[{{Baglin}(1991)}]{baglin1991}
{Baglin}, A., \emph{{Stellar seismology from space - The EVRIS experiment on
  board Mars 94}}, \emph{\solphys} \textbf{133}, 155 (1991)

\bibitem[{{Baglin} et~al.(2002)}]{baglin2002}
{Baglin}, A., et~al., \emph{{COROT: asteroseismology and planet finding}}, in
  B.~{Battrick}, F.~{Favata}, I.~W. {Roxburgh}, D.~{Galadi} (eds.) Stellar
  Structure and Habitable Planet Finding, \emph{ESA Special Publication},
  volume 485, 17--24 (2002)

\bibitem[{{Ballard} et~al.(2011)}]{ballard2011}
{Ballard}, S., et~al., \emph{{A Search for Additional Planets in Five of the
  Exoplanetary Systems Studied by the NASA EPOXI Mission}}, \emph{\apj}
  \textbf{732}, 41 (2011), \eprint{1103.0010}

\bibitem[{{Benk{\H o}} \& {Szab{\'o}}(2015)}]{benkoszabo}
{Benk{\H o}}, J.~M., {Szab{\'o}}, R., \emph{{The Blazhko Effect and Additional
  Excited Modes in RR Lyrae Stars}}, \emph{\apjl} \textbf{809}, L19 (2015),
  \eprint{1507.06814}

\bibitem[{{Benk{\H o}} et~al.(2016){Benk{\H o}}, {Szab{\'o}}, {Derekas}, \&
  {S{\'o}dor}}]{benko2016}
{Benk{\H o}}, J.~M., {Szab{\'o}}, R., {Derekas}, A., {S{\'o}dor}, {\'A}.,
  \emph{{Finest light curve details, physical parameters, and period
  fluctuations of CoRoT RR Lyrae stars}}, \emph{\mnras} \textbf{463}, 1769
  (2016), \eprint{1608.06418}

\bibitem[{{Benk{\H o}} et~al.(2010)}]{benko2010}
{Benk{\H o}}, J.~M., et~al., \emph{{Flavours of variability: 29 RR Lyrae stars
  observed with Kepler}}, \emph{\mnras} \textbf{409}, 1585 (2010),
  \eprint{1007.3928}

\bibitem[{{Benk{\H o}} et~al.(2014)}]{benko2014}
{Benk{\H o}}, J.~M., et~al., \emph{{Long-timescale Behavior of the Blazhko
  Effect from Rectified Kepler Data}}, \emph{\apjs} \textbf{213}, 31 (2014),
  \eprint{1406.5864}

\bibitem[{{Borucki}(2016)}]{borucki2016}
{Borucki}, W.~J., \emph{{KEPLER Mission: development and overview}},
  \emph{Reports on Progress in Physics} \textbf{79}, 3, 036901 (2016)

\bibitem[{{Borucki} \& {Summers}(1984)}]{borucki1984}
{Borucki}, W.~J., {Summers}, A.~L., \emph{{The photometric method of detecting
  other planetary systems}}, \emph{\icarus} \textbf{58}, 121 (1984)

\bibitem[{{Bruntt} et~al.(2008)}]{bruntt2008}
{Bruntt}, H., et~al., \emph{{Polaris the Cepheid Returns: 4.5 Years of
  Monitoring from Ground and Space}}, \emph{\apj} \textbf{683}, 433-440 (2008),
  \eprint{0804.3593}

\bibitem[{{Buchler} \& {Koll{\'a}th}(2011)}]{bk11}
{Buchler}, J.~R., {Koll{\'a}th}, Z., \emph{{On the Blazhko Effect in RR Lyrae
  Stars}}, \emph{\apj} \textbf{731}, 24 (2011), \eprint{1101.1502}

\bibitem[{{Buzasi}(2002)}]{buzasi2002}
{Buzasi}, D., \emph{{Asteroseismic Results from WIRE (invited paper)}}, in
  C.~{Aerts}, T.~R. {Bedding}, J.~{Christensen-Dalsgaard} (eds.) IAU Colloquium
  185, \emph{ASPC}, volume 259, 616 (2002)

\bibitem[{{Chadid} et~al.(2010)}]{v1127}
{Chadid}, M., et~al., \emph{{First CoRoT light curves of RR Lyrae stars.
  Complex multiplet structure and non-radial pulsation detections in V1127
  Aquilae}}, \emph{\aap} \textbf{510}, A39 (2010)

\bibitem[{{Christiansen} et~al.(2011)}]{christiansen2011}
{Christiansen}, J.~L., et~al., \emph{{System Parameters, Transit Times, and
  Secondary Eclipse Constraints of the Exoplanet Systems HAT-P-4, TrES-2,
  TrES-3, and WASP-3 from the NASA EPOXI Mission of Opportunity}}, \emph{\apj}
  \textbf{726}, 94 (2011), \eprint{1011.2229}

\bibitem[{{Clementini} et~al.(2016)}]{clementini2016}
{Clementini}, G., et~al., \emph{{Gaia Data Release 1. The Cepheid and RR Lyrae
  star pipeline and its application to the south ecliptic pole region}},
  \emph{\aap} \textbf{595}, A133 (2016), \eprint{1609.04269}

\bibitem[{{Derekas} et~al.(2012)}]{derekas2012}
{Derekas}, A., et~al., \emph{{Period and light-curve fluctuations of the Kepler
  Cepheid V1154 Cygni}}, \emph{\mnras} \textbf{425}, 1312 (2012),
  \eprint{1207.2907}

\bibitem[{{Derekas} et~al.(2017)}]{derekas2017}
{Derekas}, A., et~al., \emph{{The Kepler Cepheid V1154 Cyg revisited: light
  curve modulation and detection of granulation}}, \emph{\mnras} \textbf{464},
  1553 (2017), \eprint{1609.05398}

\bibitem[{{Dziembowski}(2016)}]{dziem}
{Dziembowski}, W.~A., \emph{{Nonradial oscillations in classical pulsating
  stars. Predictions and discoveries}}, in L.~{Szabados}, R.~{Szab\'o},
  K.~{Kinemuchi} (eds.) RRL2015: High-Precision Studies of RR Lyrae Stars,
  \emph{Comm. Konkoly Obs.}, volume 105, 23--30 (2016), \eprint{1512.03708}

\bibitem[{{Evans} et~al.(2015)}]{evans2015}
{Evans}, N.~R., et~al., \emph{{Observations of Cepheids with the MOST
  satellite: contrast between pulsation modes}}, \emph{\mnras} \textbf{446},
  4008 (2015), \eprint{1411.1730}

\bibitem[{{Gruberbauer} et~al.(2007)}]{gruberbauer}
{Gruberbauer}, M., et~al., \emph{{MOST photometry of the RRdLyrae variable
  AQLeo: two radial modes, 32 combination frequencies and beyond}},
  \emph{\mnras} \textbf{379}, 1498 (2007), \eprint{0705.4603}

\bibitem[{{Guggenberger} et~al.(2011)}]{guggenberger2011}
{Guggenberger}, E., et~al., \emph{{The CoRoT star 105288363: strong
  cycle-to-cycle changes of the Blazhko modulation}}, \emph{\mnras}
  \textbf{415}, 1577 (2011), \eprint{1104.1726}

\bibitem[{{Hick} et~al.(2007){Hick}, {Buffington}, \& {Jackson}}]{hick2007}
{Hick}, P., {Buffington}, A., {Jackson}, B.~V., \emph{A procedure for fitting
  point sources in SMEI white-light full-sky maps}, in Optical Engineering +
  Applications, \emph{Proc.SPIE}, volume 6689, 66890C (2007),
  \urlprefix\url{http://dx.doi.org/10.1117/12.734808}

\bibitem[{{Howell} et~al.(2014)}]{k2}
{Howell}, S.~B., et~al., \emph{{The K2 Mission: Characterization and Early
  Results}}, \emph{\pasp} \textbf{126}, 398 (2014), \eprint{1402.5163}

\bibitem[{{Kolenberg} et~al.(2011)}]{kolenberg2011}
{Kolenberg}, K., et~al., \emph{{Kepler photometry of the prototypical Blazhko
  star RR Lyr: an old friend seen in a new light}}, \emph{\mnras} \textbf{411},
  878 (2011), \eprint{1011.5908}

\bibitem[{{Koll\'ath}(2016)}]{kollath2016}
{Koll\'ath}, Z., \emph{{The unique dynamical system underlying RR Lyrae
  pulsations}}, in L.~{Szabados}, R.~{Szab\'o}, K.~{Kinemuchi} (eds.) RRL2015:
  High-Precision Studies of RR Lyrae Stars, \emph{Comm. Konkoly Obs.}, volume
  105, 35--38 (2016), \eprint{1601.06625}

\bibitem[{{Koll{\'a}th} et~al.(2011){Koll{\'a}th}, {Moln{\'a}r}, \&
  {Szab{\'o}}}]{kmsz11}
{Koll{\'a}th}, Z., {Moln{\'a}r}, L., {Szab{\'o}}, R., \emph{{Period-doubling
  bifurcation and high-order resonances in RR Lyrae hydrodynamical models}},
  \emph{\mnras} \textbf{414}, 1111 (2011), \eprint{1102.0157}

\bibitem[{{Kuehn} et~al.(2017){Kuehn}, {Moskalik}, \& {Drury}}]{kuehn2017}
{Kuehn}, C.~A., {Moskalik}, P., {Drury}, J.~A., \emph{RR Lyrae Stars in M4}, in
  Proceedings of Joint TASC2-KASC9-SPACEINN-HELAS8 Conference, \emph{EPJ WoC},
  volume 160, 04011 (2017)

\bibitem[{{Kurtz} et~al.(2016)}]{kurtz2016}
{Kurtz}, D.~W., et~al., \emph{{EPIC 201585823, a rare triple-mode RR Lyrae star
  discovered in K2 mission data}}, \emph{\mnras} \textbf{455}, 1237 (2016),
  \eprint{1510.03347}

\bibitem[{{Moln{\'a}r} et~al.(2012)}]{molnar2012}
{Moln{\'a}r}, L., et~al., \emph{{Nonlinear Asteroseismology of RR Lyrae}},
  \emph{\apjl} \textbf{757}, L13 (2012), \eprint{1208.4908}

\bibitem[{{Moln{\'a}r} et~al.(2015)}]{molnar2015}
{Moln{\'a}r}, L., et~al., \emph{{Pushing the Limits, Episode 2: K2 Observations
  of Extragalactic RR Lyrae Stars in the Dwarf Galaxy Leo IV}}, \emph{\apj}
  \textbf{812}, 2 (2015), \eprint{1508.05587}

\bibitem[{{Moln{\'a}r} et~al.(2017a)}]{molnar2017}
{Moln{\'a}r}, L., et~al., \emph{{The additional-mode garden of RR Lyrae
  stars}}, in Proceedings of Joint TASC2-KASC9-SPACEINN-HELAS8 Conference,
  \emph{EPJ WoC}, volume 160, 04008 (2017a), \eprint{1703.02420}

\bibitem[{{Moln{\'a}r} et~al.(2017b)}]{molnar2017b}
{Moln{\'a}r}, L., et~al., \emph{{V473 Lyr, a modulated, period-doubled Cepheid,
  and U TrA, a double-mode Cepheid, observed by MOST}}, \emph{\mnras}
  \textbf{466}, 4009 (2017b), \eprint{1612.06722}

\bibitem[{{Moskalik} \& {Ko{\l}aczkowski}(2009)}]{moskalik2009}
{Moskalik}, P., {Ko{\l}aczkowski}, Z., \emph{{Frequency analysis of Cepheids in
  the Large Magellanic Cloud: new types of classical Cepheid pulsators}},
  \emph{\mnras} \textbf{394}, 1649 (2009), \eprint{0809.0864}

\bibitem[{{Moskalik} et~al.(2015)}]{moskalik2015}
{Moskalik}, P., et~al., \emph{{Kepler photometry of RRc stars: peculiar
  double-mode pulsations and period doubling}}, \emph{\mnras} \textbf{447},
  2348 (2015), \eprint{1412.2272}

\bibitem[{{Nemec} et~al.(2011)}]{nemec2011}
{Nemec}, J.~M., et~al., \emph{{Fourier analysis of non-Blazhko ab-type RR Lyrae
  stars observed with the Kepler space telescope}}, \emph{\mnras} \textbf{417},
  1022 (2011), \eprint{1106.6120}

\bibitem[{{Nemec} et~al.(2013)}]{nemec2013}
{Nemec}, J.~M., et~al., \emph{{Metal Abundances, Radial Velocities, and Other
  Physical Characteristics for the RR Lyrae Stars in The Kepler Field}},
  \emph{\apj} \textbf{773}, 181 (2013), \eprint{1307.5820}

\bibitem[{{Netzel} et~al.(2015b){Netzel}, {Smolec}, \&
  {Dziembowski}}]{netzel2015b}
{Netzel}, H., {Smolec}, R., {Dziembowski}, W., \emph{{Discovery of a new group
  of double-periodic RR Lyrae stars in the OGLE-IV photometry}}, \emph{\mnras}
  \textbf{451}, L25 (2015b), \eprint{1504.05765}

\bibitem[{{Netzel} et~al.(2015a){Netzel}, {Smolec}, \&
  {Moskalik}}]{netzel2015a}
{Netzel}, H., {Smolec}, R., {Moskalik}, P., \emph{{Double-mode
  radial-non-radial RR Lyrae stars in the OGLE photometry of the Galactic
  bulge}}, \emph{\mnras} \textbf{447}, 1173 (2015a), \eprint{1411.3155}

\bibitem[{{Plachy} et~al.(2017a){Plachy}, {Klagyivik}, {Moln{\'a}r}, \&
  {Szab{\'o}}}]{plachy2017a}
{Plachy}, E., {Klagyivik}, P., {Moln{\'a}r}, L., {Szab{\'o}}, R., \emph{{A
  modulated RRd star observed by K2}}, in Proceedings of Joint
  TASC2-KASC9-SPACEINN-HELAS8 Conference, \emph{EPJ WoC}, volume 160, 04010
  (2017a), \eprint{1704.00288}

\bibitem[{{Plachy} et~al.(2013){Plachy}, {Koll{\'a}th}, \&
  {Moln{\'a}r}}]{plachy2013}
{Plachy}, E., {Koll{\'a}th}, Z., {Moln{\'a}r}, L., \emph{{Low-dimensional chaos
  in RR Lyrae models}}, \emph{\mnras} \textbf{433}, 3590 (2013),
  \eprint{1306.1526}

\bibitem[{{Plachy} et~al.(2014)}]{plachy2014}
{Plachy}, E., et~al., \emph{{Non-linear dynamical analysis of the Blazhko
  effect with the Kepler space telescope: the case of V783 Cyg}}, \emph{\mnras}
  \textbf{445}, 2810 (2014), \eprint{1409.4706}

\bibitem[{{Plachy} et~al.(2016)}]{plachy2016}
{Plachy}, E., et~al., \emph{{Target selection of classical pulsating variables
  for space-based photometry}}, in L.~{Szabados}, R.~{Szab\'o}, K.~{Kinemuchi}
  (eds.) RRL2015: High-Precision Studies of RR Lyrae Stars, volume 105, 19--22
  (2016), \eprint{1603.07579}

\bibitem[{{Plachy} et~al.(2017b)}]{plachy2017b}
{Plachy}, E., et~al., \emph{{First observations of W Virginis stars with K2:
  detection of period doubling}}, \emph{\mnras} \textbf{465}, 173 (2017b),
  \eprint{1610.05488}

\bibitem[{{Poretti} et~al.(2014){Poretti}, {Baglin}, \& {Weiss}}]{poretti2014}
{Poretti}, E., {Baglin}, A., {Weiss}, W.~W., \emph{{The CoRoT Discovery of a
  Unique Triple-mode Cepheid in the Galaxy}}, \emph{\apjl} \textbf{795}, L36
  (2014), \eprint{1410.8331}

\bibitem[{{Poretti} et~al.(2015)}]{poretti2015}
{Poretti}, E., et~al., \emph{{CoRoT space photometry of seven Cepheids}},
  \emph{\mnras} \textbf{454}, 849 (2015), \eprint{1508.07639}

\bibitem[{{Prudil} et~al.(2017){Prudil}, {Smolec}, {Skarka}, \&
  {Netzel}}]{prudil2017}
{Prudil}, Z., {Smolec}, R., {Skarka}, M., {Netzel}, H., \emph{{Peculiar
  double-periodic pulsation in RR Lyrae stars of the OGLE collection - II.
  Short-period stars with a dominant radial fundamental mode}}, \emph{\mnras}
  \textbf{465}, 4074 (2017), \eprint{1701.00751}

\bibitem[{{Ricker} et~al.(2014)}]{ricker2014}
{Ricker}, G.~R., et~al., \emph{{Transiting Exoplanet Survey Satellite (TESS)}},
  in Space Telescopes and Instrumentation 2014, \emph{\procspie}, volume 9143,
  914320 (2014), \eprint{1406.0151}

\bibitem[{{Riess} et~al.(2016)}]{riess2016}
{Riess}, A.~G., et~al., \emph{{A 2.4\% Determination of the Local Value of the
  Hubble Constant}}, \emph{\apj} \textbf{826}, 56 (2016), \eprint{1604.01424}

\bibitem[{{Roxburgh}(2006)}]{roxburgh2006}
{Roxburgh}, I.~W., \emph{{The Quest for a European Space Mission in Stellar
  Seismology and Planet Finding}}, in M.~{Fridlund}, A.~{Baglin}, J.~{Lochard},
  L.~{Conroy} (eds.) The CoRoT Mission Pre-Launch Status, \emph{ESA Special
  Publication}, volume 1306, 521 (2006)

\bibitem[{{Smolec} et~al.(2017)}]{smolec2017}
{Smolec}, R., et~al., \emph{{BRITE observations of classical Cepheids}}, in
  Second BRITE-Constellation Science Conference, \emph{Proc. Polish Astr.
  Soc.}, volume~5, 265 (2017), \eprint{1612.02970}

\bibitem[{{S\'odor} et~al.(2006)}]{sodor2006}
{S\'odor}, A., et~al., \emph{{UZ UMa: An RRab star with double-periodic
  modulation}}, \emph{Information Bulletin on Variable Stars} \textbf{5705}
  (2006)

\bibitem[{{Szab{\'o}} et~al.(2010)}]{szabo2010}
{Szab{\'o}}, R., et~al., \emph{{Does Kepler unveil the mystery of the Blazhko
  effect? First detection of period doubling in Kepler Blazhko RR Lyrae
  stars}}, \emph{\mnras} \textbf{409}, 1244 (2010), \eprint{1007.3404}

\bibitem[{{Szab{\'o}} et~al.(2011)}]{szabo2011}
{Szab{\'o}}, R., et~al., \emph{{Cepheid investigations using the Kepler space
  telescope}}, \emph{\mnras} \textbf{413}, 2709 (2011), \eprint{1101.2443}

\bibitem[{{Szab{\'o}} et~al.(2014)}]{szabo2014}
{Szab{\'o}}, R., et~al., \emph{{Revisiting CoRoT RR Lyrae stars: detection of
  period doubling and temporal variation of additional frequencies}},
  \emph{\aap} \textbf{570}, A100 (2014), \eprint{1408.0653}

\bibitem[{{Walker} et~al.(2003)}]{walker2003}
{Walker}, G., et~al., \emph{{The MOST Asteroseismology Mission: Ultraprecise
  Photometry from Space}}, \emph{\pasp} \textbf{115}, 1023 (2003)

\end{thebibliography}

\end{document}